\documentclass[doublecol]{epl2} 

\usepackage{amsmath,amssymb,amsfonts}
\usepackage{graphicx}
\usepackage{color}

\DeclareMathOperator{\csch}{csch}

\DeclareMathOperator{\La}{\mathcal{L}}

\title{Tunnelling of the 3rd kind: A test of the effective non-locality of quantum field theory}
\shorttitle{T3K: Effective Non-locality in QFT} 

\author{Simon A. Gardiner\inst{1} \and Holger Gies \inst{2,3} \and Joerg Jaeckel\inst{4} \and Chris J. Wallace\inst{1}}
\shortauthor{S.~A.~Gardiner \etal}

\institute{
  \small
  \inst{1} Department of Physics, Durham University, Durham DH1 3LE, United Kingdom\\
  \inst{2} Theoretisch-Physikalisches Institut, Friedrich-Schiller-Universit\"{a}t Jena,
Max-Wien-Platz 1, D-07743 Jena, Germany\\
  \inst{3} Helmholtz-Institut Jena, Fr\"obelstieg 3, D-07743 Jena, Germany\\
  \inst{4} Institut f\"{u}r Theoretische Physik, Universit\"{a}t Heidelberg, Heidelberg, Germany
  \normalsize    \fontsize{10pt}{12pt}\selectfont
}

\abstract{\noindent Integrating out virtual quantum fluctuations in an originally local quantum field theory results in an effective theory which is non-local. In this Letter we argue that tunnelling of the 3rd kind --- where particles traverse a barrier by splitting into a pair of virtual particles which recombine only after a finite distance --- provides a direct test of this non-locality.  We sketch a quantum-optical setup to test this effect, and investigate observable effects in a simple toy model.
}

\begin{document}

\maketitle

\section{Introduction}
\begin{figure}[t]
\centerline{\includegraphics[width=7.8cm]{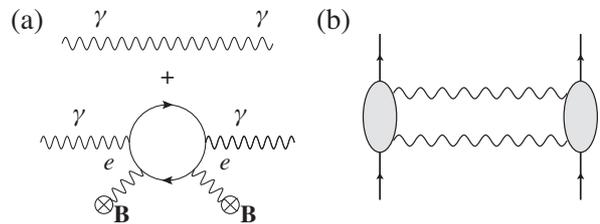}}
\caption{(a) Contributions to the Euler--Heisenberg Lagrangian. The first term is the classical part and the second corresponds to the effect of virtual particles in an external magnetic field~$\mathbf{B}$. (b) Leading order contribution to the Casimir--Polder force between two polarizable molecules. The interaction of the photon with the molecule is indicated by the grey ellipse.
}
\label{fig:euler}
\end{figure}

The existence of virtual particles --- field configurations which do not fulfil the classical equations of motion --- is an inherent feature of quantum fields. Such particles lead to a host of interesting and novel effects which find a natural description in the language of effective field theories. One of the earliest quantum field theory calculations that included the effect of virtual particles was the derivation of the Euler--Heisenberg Lagrangian~\cite{Heisenberg:1935qt,Dittrich:2000zu,Dunne:2012vv}. The theory included higher dimensional operators to characterize the effects of virtual fermion loops in light propagation, allowing photons an effective self-interaction [see Fig.~\ref{fig:euler}(a)].  The Uehling potential~\cite{Uehling:1935uj} is a correction to the electrostatic potential caused by an electron loop similar to that shown in Fig.~\ref{fig:euler}(a) (but without the background field) and may be measured as a contribution to the Lamb shift of the hydrogen spectrum, itself an effect of virtual particles.  Another example arises in calculating the van der Waals force between two polarizable molecules. Taking into account retardation effects in the calculation yields the Casimir--Polder force~\cite{Casmir:1947hx,Jaffe:2005vp} [see Fig.~\ref{fig:euler}(b)].  In each of these examples the effects of virtual particles are manifest either in the form of \textit{local\/} higher dimensional operators (Euler--Heisenberg), or in generating contributions to effective potentials between two particles (Uehling potential and Casimir--Polder force).  In this Letter we describe an effect directly demonstrating that the presence of loops also leads to an effective \textit{non-locality in the propagation of particles} and discuss how this effect could be realized in low energy experimental configurations. This effective non-locality in the propagation is an aspect of quantum field theory which, to our knowledge, has not been tested experimentally.

\section{Nonlocality in Effective Field Theories}
\begin{figure}[t]
\centerline{\includegraphics[width=7.8cm]{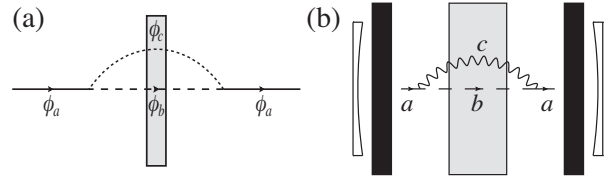}}
\caption{Schematic processes for tunnelling of the 3rd kind. 
(a) The incoming real particle $\phi_{a}$ splits into two virtual particles $\phi_{b}$ and $\phi_{c}$ which do not interact with the barrier (grey shaded). After traversing the barrier they recombine into $\phi_{a}$. In this way the particle $\phi_{a}$ can ``tunnel'' through the barrier. 
(b) The incoming atom $a$ emits a virtual photon $c$ and enters an excited state $b$. After traversing the barrier they recombine and return to the original state.  The mirrors on either side represent an optical cavity used to enhance the coupling of photons to the atom. The black walls denote potential barriers blocking both $a$ and $b$, confining the atom and preventing contact with the mirrors.}
  \label{toyfig}
\end{figure}

The starting point of most quantum field theories (QFTs) is a classical \textit{local\/} field theory. The action $S\equiv \int d^{4}x \La$ arises as an integral over a local Lagrangian density in which all fields are evaluated at the same point, and typically no more than two derivatives appear. As an example, we consider
an interaction of three scalar fields $\phi_{a,b,c}$ (to connect with what follows we take $\phi_{a,b}$ complex and $\phi_{c}$ real),
\begin{equation}
\begin{split}
S_{\rm int} =& -\int d^{4}x
\frac{\lambda}{\hbar^2 c^2}\phi_{a}(x)\phi^{*}_{b}(x)\phi_{c}(x)+{\rm c.c.} \,.
\end{split}
\label{toylag}
\end{equation}
Together with standard kinetic terms and possible couplings to external potentials,
the action described by Eq.~\eqref{toylag} is entirely local. Consequently, the equation of motion for a field at a given space-time point contains only the field and its derivatives at that point. The future evolution is entirely determined by the field value at this point and its immediate surroundings.  Alternatively, upon second quantization this becomes a theory of point-like particles with contact interactions. Quantum fluctuations allow for the production of virtual particles that briefly exist and then vanish. For example, with \eqref{toylag} a particle of type $\phi_{a}$ can split into a transient $\phi_{b}$ and $\phi_{c}$ pair, which subsequently recombine into the original $\phi_{a}$, as shown in Fig.~\ref{toyfig}(a). Although originating entirely from local interactions the effect on the propagation of the \textit{real\/} particle $\phi_{a}$ is \textit{non-local}; it seems to disappear at some point, only to reappear elsewhere.

A convenient and illuminating way to account for virtual particle effects is to use an appropriate effective action. The effects of virtual fluctuations are encoded into modified propagation and interaction terms (for simplicity we consider the theory in $1+1$ dimensions and in the non-relativistic limit):
\begin{multline}
\label{efflag}
S_{\textrm{eff}}= \iiiint dt dt' dx dx'
\phi_{a}^{*}(x',t')
\bigg\{\delta(x-x')\delta(t-t')
\\
\times\bigg[
i\hbar\partial_{t} + 
\frac{\hbar^{2}}{2m}\partial_{x}^{2}
- V_{a}(x)
\bigg]
-\Pi(x,x',t-t')\bigg\}\phi_{a}(x,t),
\end{multline}
plus terms involving $\phi_{b}$ and $\phi_{c}$ that we ignore in the following. The terms proportional to $\delta(x-x^{\prime})\delta(t-t^{\prime})$ are the non-relativistic local kinetic and potential terms. The next term accounts for the effects of virtual particles [see, e.g., Fig.~\ref{toyfig}(a)]. Assuming time-independent potentials, $\Pi$ depends only on the time difference $t-t^{\prime}$. In particular, when $\Pi(x,x^{\prime},t-t')$ is not proportional to $\delta(x-x^{\prime})$, this describes a spatial non-locality in the effective propagation of $\phi_{a}$. 
The equation of motion generated by Eq.~\eqref{efflag},
\begin{equation}
\label{sch}
\begin{split}
i\hbar \partial_{t}\phi_{a}(x,t)=&-\frac{\hbar^2}{2 m}\partial^{2}_{x}\phi_{a}(x,t)+V(x)\phi_{a}(x,t)
\\ &+\iint
dx^{\prime}dt^{\prime}\Pi(x,x^{\prime},t-t^{\prime})\phi_{a}(x^{\prime},t^{\prime}),
\end{split}
\end{equation}
can be viewed as an effective Schr\"odinger equation for $\phi_{a}(x,t)$. 
In contrast to the standard Schr\"odinger equation, the evolution of the field $\phi_{a}(x,t)$ at point $x$ depends not only on its value at $x$ and its immediate surroundings, but also on the value of the field at \emph{finite} distances, as explicit in the second line of Eq.~\eqref{sch}.

In this Letter we present a direct way to demonstrate this effective non-locality, as sketched in Fig.~\ref{toyfig}(a). We start with a particle of type $\phi_{a}$, which encounters a potential barrier (grey). In ordinary quantum mechanics, if the barrier is of finite height and width there is a non-zero tunnelling probability; if the barrier is infinitely high this probability is zero. In quantum field theory (which we distinguish from quantum mechanics by its permitting non-particle-number-conserving processes), the particle may also split into a pair of virtual particles $\phi_{b}$ and $\phi_{c}$~\cite{Gies:2009wx}. If these particles interact only very weakly with the barrier, they can traverse it and then recombine into a real $\phi_{a}$ particle; the real particle $\phi_{a}$ appears to have vanished on one side of the barrier only to reemerge on the other --- a clearly non-local effect which we will refer to in the following as ``tunnelling of the 3rd kind'' (T3K)\footnote{This is in contrast to regular quantum tunnelling and a second kind, where a particle traverses a barrier via classical conversion to another real particle species that can pass the barrier. Such a barrier crossing (by conversion to another species) is the operating principle of so-called ``light-shining-through-walls'' experiments \cite{Redondo:2010dp}.} \cite{Gies:2009wx}.

We emphasize that this effect is purely due to the existence of quantum fields, in the sense of arising explicitly through virtual particle creation, and has no classical or even single-particle quantum-mechanical analogue. This type of non-locality in an effective description of particle propagation should not be confused with a non-local correlation in the context of an EPR experiment (i.e. the statement that the predictions of non-relativistic quantum mechanics cannot be explained by a locally realistic theory, and hence violate Bell's inequalities). Here the effect appears only at loop-level, and is really induced explicitly by the temporary violation of particle number that is present in QFT but not in non-relativistic quantum mechanics. As such, the effects, though both in some sense non-local, are totally independent.

We propose a toy model setup with atoms within an electromagnetic resonator that realizes the essential ingredients necessary for T3K, thereby allowing for a direct test of the effective non-locality of quantum field theory. We then study T3K in this toy model and calculate the transition probability. 

In the Standard Model the only particles suitable as intermediate states for T3K are neutrinos; the probability, however, is prohibitively small~\cite{Gies:2009wx}.  A search for T3K can in fact be sensitive to as yet undiscovered beyond Standard Model particles.  Such searches may even improve on bounds from cosmology~\cite{Dobrich:2012sw,Dobrich:2012jd}.  Fortunately,  it is not necessary to use elementary particles to test T3K and therefore the corresponding non-locality; we can instead use bound states such as atoms or molecules, which can realize all  necessary features. The effect is precisely the same in each case --- the low energy system, however, allows us to make this effect directly visible in a more feasible experimental set up.

\section{Toy Quantum-Optical Configuration}
Consider specifically the process depicted in Fig.~\ref{toyfig}(b): an atom emits a virtual photon and enters a (virtual) excited state; the photon is subsequently reabsorbed and the atom returns to its original state. This is a loop process closely analogous to that depicted in Fig.~\ref{toyfig}(a), and represents a quantum field theory correction to the propagation of the atom, which is exactly the type of process we expect for T3K and for generating the effective non-locality of quantum field theory.

The only missing ingredient is for the excited atom and the virtual photon to be able to penetrate the barrier, but the atom in its original state being unable to do so. This may be realized within cold atom systems, where atoms are commonly trapped by potentials generated by electromagnetic fields~\cite{trapping1,trapping2,trapping3}; such potentials may be used to produce a barrier. Indeed such systems of barriers have been used, e.g. in~\cite{oberthaler}, to demonstrate ordinary tunnelling of atoms. Potentials of this type can be highly sensitive to the internal state of the atom; accordingly, one may arrange for a situation such that the ground-state atom is blocked but a virtual photon and a suitable excited state are not. 

Finally, we consider two further refinements of the setup that simplify the calculation and may also aid practical realization. Firstly, we place the whole system within an optical or microwave cavity that allows for only one relevant resonance mode~\cite{cavity}. This can also increase the coupling to the atom, thereby magnifying the effect.  Secondly, as the atoms (both ground and excited states) should preferably not hit the cavity mirrors, we include two further potential barriers to confine the atoms within the cavity. We therefore have two potential wells separated by a central barrier, which is, however, transparent to the excited state [see Fig.~\ref{toyfig}(b)]. The question of T3K can now be rephrased as: if we start at time $t=0$ with an atom in the ground state, entirely localized on the left hand side, what is the probability to find it at a later time $t>0$ in the right hand side potential well? 

Let us first link the tunnelling rate to the non-local term in our effective Lagrangian~\eqref{efflag}. We use an eigenbasis expansion of the field $\phi_{a}$ to explicitly separate time and position dependence, expanding $\phi_{a}(x,t) = \sum_{n} c^{(a)}_{n}(t) \varphi^{(a)}_{n}(x)$ in energy eigenstates $\varphi^{(a)}_{n}(x)$ of the unperturbed system, as defined by
\begin{equation}
\left[-\frac{\hbar^2}{2m} \partial_x^2 + V_{a}(x)\right] 
\varphi_n^{(a)} (x) = E^{(a)}_n \varphi_n^{(a)}(x).
\label{unperturbedeigenbasis}
\end{equation}
Inserting the eigenbasis expansion into \eqref{efflag} and performing the spatial integrals, we complete the transformation from position to energy space, leaving us with
\begin{equation}
\begin{split}
S_{\textrm{eff}} =& \sum_{n} \int dt
\left[c^{(a)*}_{n}(t)\left(i\hbar\frac{d}{dt} - E^{(a)}_{n} \right) c^{(a)}_{n}(t) \right] 
\\
&- \sum_{n,n'} \iint dt dt^{\prime} \left[c^{(a)*}_{n}(t)\Pi_{n,n'} \delta(t-t^{\prime})c^{(a)}_{n'}(t^{\prime})\right],
\end{split}
\label{actiondiracdelta}
\end{equation}
where
$\Pi_{n,n^{\prime}}(t-t') \equiv \iint dx dx^{\prime} \varphi^{(a)*}_{n}(x)\Pi(x,x^{\prime},t-t')\varphi^{(a)}_{n^{\prime}}(x^{\prime}).
$
We have also made a simplifying assumption, appropriate for our system. If the intermediate (virtual) two-particle states have very different energies from the initial and final one-particle states, the former are much shorter-lived, and $\Pi_{n,n'}(t-t') \approx \Pi_{n,n'}\delta(t-t^{\prime})$. This approximation requires us to stay sufficiently far from resonances, i.e., we assume that the $\phi_c$ excitations are sufficiently detuned with respect to the gap between the excited $\phi_b$ states and the $\phi_a$ ground state. This will also be justified \emph{a posteriori} by the agreement between the dominant parts of the result obtained from the Lagrangian formalism and our Hamiltonian calculation (which makes no further assumption than being far from resonance).

We determine the equations of motion by taking a functional derivative of \eqref{actiondiracdelta} with respect to $c^{(a)*}_{n}(t)$:
\begin{equation}
0=\frac{\delta S_{\textrm{eff}}}{\delta c^{(a)*}_{n}(t)}=\left(i\hbar\frac{d}{dt}-E^{(a)}_{n}\right)c^{(a)}_{n}(t)-\sum_{n'}\Pi_{n,n'}c^{(a)}_{n'}(t).
\end{equation}
For simplicity we describe $\phi_{a}$ using only the two lowest energy eigenfunctions of \eqref{unperturbedeigenbasis}; if we further assume $V_{a}$ to be symmetric about the central barrier between the two potential wells, the ground  and first-excited state energy eigenfunctions are symmetric and antisymmetric, respectively.  Referring to the associated amplitudes and eigenenergies as $c_{S,A}$ and $E_{S,A}$, the equations of motion reduce to a $2\times2$ matrix equation:
\begin{equation}
\begin{split}
i\hbar \frac{d}{dt} 
    \left( \begin{array}{c}
    c_{S}(t)  \\
     c_{A}(t)
     \end{array}\right)
      =&
\left( \begin{array}{cc}
    E_{S} & 0  \\
     0 & E_{A}
     \end{array}\right)
     \left( \begin{array}{c}
    c_{S}(t)  \\
     c_{A}(t)
     \end{array}\right)     \\
    &+ \left( \begin{array}{cc}
    \Pi_{SS} & 0  \\
     0 & \Pi_{AA}
     \end{array}\right)
     \left( \begin{array}{c}
    c_{S}(t)  \\
     c_{A}(t)
     \end{array}\right).
\end{split}
\end{equation}
Parity symmetry requires $\Pi_{AS}=\Pi_{SA}=0$, and letting the barrier between the two wells tend to infinity we may set $ E\equiv E_{S}=E_{A}$;
$\Pi_{SS}$ and $\Pi_{AA}$ then describe energy shifts caused by the creation and annihilation of virtual particles. 
As is well known, the energy shift $\delta E=\Pi_{AA}-\Pi_{SS}$ determines the leading order tunnelling rate between the two wells, which in our setup 
occurs solely non-locally via virtual intermediate states of the fluctuating fields. Defining ``left'' and ``right'' amplitudes through $c_{L}\equiv (c_{S}+c_{A})/\sqrt{2}$, $c_{R}\equiv (c_{S}-c_{A})/\sqrt{2}$, and solving for the initial condition $c_{L}(0)=1$, $c_{R}(0)=0$, we find
$c_{R}(t)=[e^{-i (E+\Pi_{SS})t/\hbar}-e^{-i(E+\Pi_{AA})t/\hbar}]/2$,
and for the tunnelling probability
\begin{equation}
\label{pitunnelling}
P_{\textrm{T3K}}(t)=|c_{R}(t)|^2=\sin^{2}([\Pi_{AA}-\Pi_{SS}]t/2\hbar).
\end{equation}
This is of course exactly the field-theory analogue of Rabi oscillations with the characteristic tunnelling rate $\sim \delta E/\hbar$. 

\section{Hamiltonian Description}
The features of the toy atomic system we have described may also be determined within a Hamiltonian description. Both descriptions have their advantages. The effective field theory description, derived from QFT, makes the non-locality manifest (Eq.~\eqref{sch}), whereas the Hamiltonian formalism is more convenient for calculating the tunnelling probability.

Let us map the field-theory formulation onto a toy model with (countably infinite) degrees of freedom.  We associate the field $\phi_b$ with atoms of type $b$ having internal energy $\hbar\Omega_b$. These are subject to the exterior trapping potentials
[black walls in Fig.~\ref{toyfig}(b)] which we
model by idealized potential walls $V_{b}(x)$ separated by a distance $2\ell+d$,
\begin{equation}
V_{b}(x)=  \left\{\begin{array}{ccc}
    0&{\rm for}& -\ell-d/2\leq x\leq \ell+d/2, \\
    \infty & {\rm for} & |x|>\ell+d/2.\\ 
  \end{array}\right.
\end{equation}
Similarly, the field $\phi_a$ is associated with atoms of type $a$ with internal energy $\hbar \Omega_a$. In addition to the exterior potential walls, the atoms $a$ are subject to the central barrier [grey region in Fig.~\ref{toyfig}(b)] of thickness $d$, modelled by
\begin{equation}
U(x)=  \left\{\begin{array}{ccc}
    \infty &{\rm for}& -d/2<x<d/2, \\
    0 & {\rm for} & |x|\geq d/2.\\ 
  \end{array}\right.
\end{equation}
Hence, the total potential experienced by atoms in state $a$ is $V_{a}(x) = V_{b}(x) + U(x)$. The corresponding bare eigenenergies for atoms in internal states $a$, $b$ are therefore given by $\hbar\omega^{(a)}_{j}$, $\hbar\omega^{(b)}_{j}$, where 
\begin{align}
\omega^{(a)}_{j}=&\Omega_a+
\frac{\hbar \pi^2 j^2}{2m\ell^2},
&
\omega^{(b)}_{j}=&\Omega_b+ \frac{\hbar
  \pi^2 j^2}{2m(2\ell+d)^2},
\label{eq:Eb}
\end{align}
and the $\omega^{(a)}_{j}$ are all doubly degenerate (corresponding to energy levels in the left and right infinite potential wells). For simplicity, we again consider only the two (lowest energy) $j=1$ spatial modes for the $a$ atoms, which we call $\varphi^{(a)}_{L}(x)$, $\varphi^{(a)}_{R}(x)$, also setting $\omega_{a}\equiv\omega^{(a)}_{1}$; we call the spatial modes for the $b$ atoms $\varphi^{(b)}_{j}(x)$. We further assume that only a single relevant cavity mode couples the $a$ and $b$ atoms, contributing $\hbar\omega_{c}\hat{c}^{\dagger}\hat{c}$ to the system Hamiltonian, where $\hat{c}^{\dagger}$ creates a photon in the cavity mode function $C(x)$, and $\omega_{c}$ is the corresponding frequency. We consider a standard dipole coupling between the atoms and the cavity photons, and, formulated as a non-relativistic quantum field theory, the complete model Hamiltonian is then
\begin{multline}
\hat{H}= \sum_{\sigma=L,R} \hbar\omega_a \hat{a}_\sigma^\dagger \hat{a}_\sigma +
\sum_{j=1}^\infty \hbar\omega^{(b)}_{j} \hat{b}_j^\dagger \hat{b}_j + \hbar\omega_c \hat{c}^\dagger \hat{c}
\\
+ \sum_{\sigma=L,R}\sum_{j=1}^{\infty} \hbar g_{\sigma j} 
(\hat{c} + \hat{c}^\dagger)
(\hat{a}_\sigma^\dagger \hat{b}_j + \hat{b}_j^\dagger \hat{a}_\sigma) ,
\label{eq:2ndQH}
\end{multline}
where $\hat{a}_\sigma^\dagger $ creates an atom in internal state $a$ and spatial mode $\varphi^{(a)}_{\sigma}(x)$, and $\hat{b}_j^\dagger$ creates an atom in internal state $b$ and spatial mode $\varphi^{(b)}_{j}(x)$, and the couplings are given by the integral
\vspace*{-0.3cm}
\begin{equation}
g_{\sigma j} = g \int dx \varphi^{(a)}_\sigma(x) \varphi^{(b)}_j(x)
C(x),
\label{eq:modecouplings}
\end{equation}
together with a coupling parameter $g$. The phase conventions are chosen such that the mode couplings $g_{\sigma j}$ are real.  Note that we have not carried out any rotating wave approximation (RWA) (e.g. \cite{rwa1}), whereby terms proportional to $\hat{c} \hat{a}_\sigma^\dagger\hat{b}_j$, $\hat{c}^\dagger \hat{b}_j^\dagger \hat{a}_\sigma$ would be discarded. The effects of most interest to us in fact manifestly arise from these beyond-RWA terms \cite{rwa2,rwa3}. 

We will consider just a single atom, which may be in either of the two internal states $a$, $b$.  In this case we may restrict ourselves to a basis $\{|a,\sigma,n\rangle, |b,j,n\rangle \}$, where $\sigma \in \{L,R\}$ and $j\in\{1,2,3,\ldots \}$ denote the spatial mode, and $n \in \{0,1,2,\ldots\}$ the number of cavity photons. We can therefore conclude that T3K has taken place if an initially pure $|a,L,0\rangle$ state is observed in an $|a,R,0\rangle$ state; to avoid real transitions we consider the situation
where there are no photons in the initial and final states. The
probability for tunnelling of the 3rd kind is then
\begin{equation}
P_{\textrm{T3K}}(t)=|\langle a,R,0|\exp(-i\hat{H}t/\hbar)|a,L,0\rangle|^{2}.
\end{equation} 

An illustrative, although over-naive first approximation for $P_{\textrm{T3K}}$ may be obtained using a basis containing states in only the lowest spatial modes and with at most one photon.
The subspace 
$\{
|a,L,0\rangle,
|b,1,1\rangle,
|a,R,0\rangle
\}$ 
decouples, and we may represent the Hamiltonian in this basis by  
\begin{equation}
\label{hammatrix}
\hat{H}= \hbar \left( \begin{array}{ccc}
    \omega_{a} & \tilde{g} & 0  \\ 
    \tilde{g} & \omega_{b}+\omega_{c} & \tilde{g}\\ 
     0 &\tilde{g} & \omega_{a}
  \end{array}\right),
\end{equation}
where $\omega_b\equiv \omega^{(b)}_{1}$ and $\omega_a\equiv \omega^{(a)}_{1}$ are given in Eq.~\eqref{eq:Eb}, and $\tilde{g}=g_{L1}= g_{R1}$ is defined in Eq.~\eqref{eq:modecouplings}.  It is straightforward to solve the time evolution in this case.  To lowest order in $\tilde{g}$
\begin{equation}
P_{\textrm{T3K}}(t) =
\left|\sin(\tilde{g}^{2}t/\delta)
+2(\tilde{g}/\delta)^{2}e^{-i\delta t/2}\sin(\delta t/2)\right|^{2},
\label{estimate}
\end{equation}
where $\delta = \omega_{b}+\omega_{c}-\omega_{a}$. Assuming $|\tilde{g}| \ll |\delta|$, there is a slow oscillation with frequency $2\tilde{g}^{2}/\delta$, and a rapid oscillation with frequency $\delta$, the amplitude of which is suppressed by $(\tilde{g}/\delta)^{2}$. For small $t$ both parts contribute to a comparable degree, but for large $t$ the first term in \eqref{estimate} dominates.  We also see that neglecting the second term\footnote{Our present approximation in the effective action calculation assumes that the interaction is localized in time, which is not true due to the confining potentials in our problem.} in \eqref{estimate} yields the same time-dependence as the result \eqref{pitunnelling} from our Lagrangian formulation, if we identify $\Pi_{AA}- \Pi_{SS}= 2\tilde{g}^2/ \delta$ (justifying our earlier assumption of time-locality).  Indeed one may readily verify that $2\tilde{g}^2/ \delta$ corresponds to the energy difference between the symmetric and antisymmetric states calculated in second order perturbation theory.
Generally, however,  the energy required to transfer from $|a,\sigma,0\rangle$ to $|b,j,1\rangle$ ($\sim \hbar\delta$) is much larger than the energy splittings between the spatial modes of $b$ atoms [$= \pi^2\hbar^2/2m (2\ell+d)^2$ --- see \eqref{eq:Eb}]. Realistically, we should therefore also include all such spatial modes where $j>1$.  We also note that, if the intermediate state $|b,j,1\rangle$ has a higher energy than the original state $|a, L,0\rangle$, i.e., $\delta>0$, then one would expect from the Heisenberg uncertainty principle that the tunnelling rate is exponentially suppressed with increasing energy difference and with increasing wall thickness. This is not exhibited by \eqref{estimate}. 

Including higher order spatial modes of the virtual $b$ state 
causes the matrix \eqref{hammatrix} to become infinite dimensional, but we can nevertheless obtain a result in second order perturbation theory.  We consider a simplified situation where the cavity mode $C(x)$ is sufficiently slowly varying for us to take $\tilde{g}_{0} \equiv gC(x)$ in \eqref{eq:modecouplings} to be constant. Second order perturbation theory then yields, for $\delta E =  \Pi_{AA} - \Pi_{SS}$,
\begin{equation}
\delta E
=
\frac{
8\tilde{g}_{0}^2 m\xi\ell 
\sinh^{2}(\ell/\xi)
\csch([2\ell+d]/\xi)}
{\pi^2 (1+\ell^2/\xi^2 \pi^2)^2},
\label{energydiff}
\end{equation}
where $\xi = \sqrt{\hbar/2m|\Delta|}$ is a characteristic length scale, and 
$\Delta=\Omega_{b}+\omega_{c}-\omega_a$. For $\Delta>0$ and in the limit $\ell \gg \xi$, \eqref{energydiff} reduces to
\begin{equation}
\delta E \approx  \epsilon \exp(-d/\xi),
\end{equation} 
which exhibits the expected exponential behaviour in the wall thickness. Here
\begin{equation}
\epsilon =
2\pi^2\frac{\hbar\tilde{g}_{0}^2}{\Delta} \,\left(\frac{\xi}{\ell}\right)^{3}
\label{energydiffreduced}
\end{equation}
is a characteristic energy scale, which represents an upper bound to $\delta E$.

\section{Experimental Considerations}
The exponential mass dependence encoded in $\xi$ makes an experiment challenging.
From Eq.~\eqref{energydiffreduced} it is clear the length scale of the effective nonlocality, i.e. the maximum barrier width, $d = \xi \ln(\epsilon/\delta E)$, is primarily determined by $\xi$, having only a weak logarithmic dependence on the ratio $\epsilon/\delta E$.  The requirement that the tunnelling rate $\delta E/\hbar$ be significantly greater than the cavity decay rate $\kappa$ then directly implies
\begin{equation}
d \ll  \xi \ln \left( 
\frac{\epsilon}{\hbar \kappa}
\right).
\end{equation}

As an example let us consider superconducting microwave cavities and circular Rydberg atoms~\cite{tweezers}. The cavity setup must be cryogenically cooled (hence superconducting) to eliminate the effects of black-body radiation, something which is not an issue in the optical domain. 
Two numbers are crucial: the typical decay length $\xi$ and the typical transition frequency $\epsilon/\hbar$.  The single photon Rabi frequency corresponds to $\tilde{g}_{0}$ in our calculation. For the type of system we are looking at, an achievable number is $\Omega/2\pi = 50$~kHz, for a 2-level transition frequency of 51.1 GHz. 
The equivalent cavity decay rate is $\kappa/2\pi = 7.7$~Hz, and the atomic lifetimes are of the order of seconds~\cite{tweezers}.
In this case $\xi \sim 30\,{\rm pm}$ and $\epsilon/\hbar \sim 2\pi^2 \tilde{g}^{2}_{0}/\Delta \sim 6\,{\rm Hz}$, where we have optimistically assumed $\ell\sim\xi$. While the latter is of the order of magnitude of the decay rate, $\xi$ is too small to be feasible in an optical setup.
We note that within an optical cavity configuration, state-dependent potentials exploiting the atom's resolved magnetic substructure would be rather more straightforward, but the available parameter regime is otherwise more challenging.

Experimentally, the situation where the energy of internal state $a$ is
higher than that of state $b$ with an additional photon is more promising. We can determine $\delta E$ perturbatively in the same way as above, which amounts to replacing all occurrences of $\Delta$ in
\eqref{energydiff} with $-|\Delta|$ (hence $\xi$ is replaced with $i\xi$). In this way we obtain
\begin{equation}
\label{negenergydiff}
\delta E
=
\frac{
8\tilde{g}_{0}^2m\xi\ell 
\sin^{2}(\ell/\xi)
\csc([2\ell+d]/\xi)
}{ \pi^2(1-\ell^2/\xi^2 \pi^2)^2}.
\end{equation}
Note that in this scenario the energy difference
no longer decays exponentially with the wall
thickness, making the system potentially realizable.  Fundamentally
this is the same process as in the $\Delta>0$ case; although a real
decay with subsequent absorption is energetically possible, the cavity
only allows for ``wrong frequency photons''\footnote{This is analogous to
  the situation in~\cite{Gies:2009wx} where for $\omega>2m$ a
  decay/absorption is energetically possible but forbidden by momentum
  conservation.} such that $\Delta\neq 0$. Hence the intermediate
state has a different energy from the initial state and is virtual in
this sense.

\section{Conclusions}
We have examined the realization of tunnelling of the 3rd kind, an effect first proposed in a high-energy context \cite{Gies:2009wx}, in a toy
quantum-optical system. In particular, we have described a setup where atoms transit from
one side to the other of an (infinite) potential double well via an intermediate virtual 2-particle (atom-photon) state. 
From the point of view of the original atom this transition is non-local, with the atom ``disappearing'' on one side and emerging on the other.
The toy model realises all the features of this tunnelling effect, with the advantage over a high-energy system of being potentially realizable.
In this Letter we have considered an experimental situation far away from resonances. The resonant case may exhibit additional interesting features; in particular, the approximate time-locality may not be valid, leading to further interesting effects. Moreover, the resonant case could simulate a highly topical ``light-shining-through-walls'' experiment~\cite{Redondo:2010dp} used to search for new light particles beyond the Standard Model, such as axions and hidden photons~\cite{Jaeckel:2010ni}.   

\acknowledgments
We would like to thank Jens Braun, Valya Khoze and Frank Krauss for useful discussions. SAG thanks the UK EPSRC 
(Grant No.\ EP/G056781/1) for support. HG acknowledges support by the DFG under grants SFB-TR18 and
GI~328/5-1 (Heisenberg program).

\end{document}